\documentclass[aps,prb,preprint,groupedaddress,showpacs]{revtex4-1}
\usepackage{graphicx}
\usepackage{subfig} 
\usepackage{epstopdf}
\usepackage{amssymb}

\usepackage{amsmath}

\usepackage{bm}
\usepackage{epstopdf}
\usepackage{amssymb}
\usepackage{xcolor}
\usepackage[normalem]{ulem}
\usepackage{bm}
\begin{document}
\title{Monte Carlo Simulations of ABC Stacked Kagome Lattice Thin Films}


\author{H.V. Yerzhakov, M.L. Plumer, and J.P. Whitehead}
\address{Department of Physics and Physical Oceanography, Memorial University of Newfoundland, St. John's, Newfoundland, Canada, A1B 3X7}

\date{\today}

\begin{abstract}
Properties of thin films of geometrically frustrated  ABC stacked antiferromagnetic kagome layers are examined using Metropolis Monte Carlo simulations. The impact of having an easy-axis anisotropy on the surface layers and cubic anisotropy in the interior layers is explored. The spin structure at the surface is shown to be different from that of the bulk 3D fcc system, where surface axial anisotropy tends to align spins along the surface [111] normal axis. This alignment tendency then propagates only weakly to the interior layers through exchange coupling. Results are shown for the specific heat, magnetization and sub-lattice order parameters for both surface and interior spins in three and six layer films as a function of increasing axial surface anisotropy. Relevance to the exchange bias phenomenon in IrMn$_3$ thin films is discussed. 
\end{abstract}




\maketitle
\section{Introduction}

Investigations of the antiferromagnetic (AF) kagome lattice continues to reveal unusual classical and quantum spin structures \cite{gorohovsky2015,carrasquilla2015} largely due to the macroscopic degeneracy associated with the four nearest neighbors (NN) of corner sharing triangles \cite{chalker1992,harris1992,schnabel2012}.  With only Heisenberg (or XY) exchange interactions, q=0 or $\sqrt{3} \times \sqrt{3}$ magnetic order characterized by planar 120$^0$ spin structures are among the degenerate ground states of the 2D lattice. Three-dimensional structures composed of weakly coupled kagome layers having rhomohedral symmetry or distorted hyperkagome lattice structures have also been studied \cite{mendels2011,zhitomirsky2008}. 

\begin{figure}[ht!bp]
\begin{center}
\includegraphics[width=0.5\textwidth]{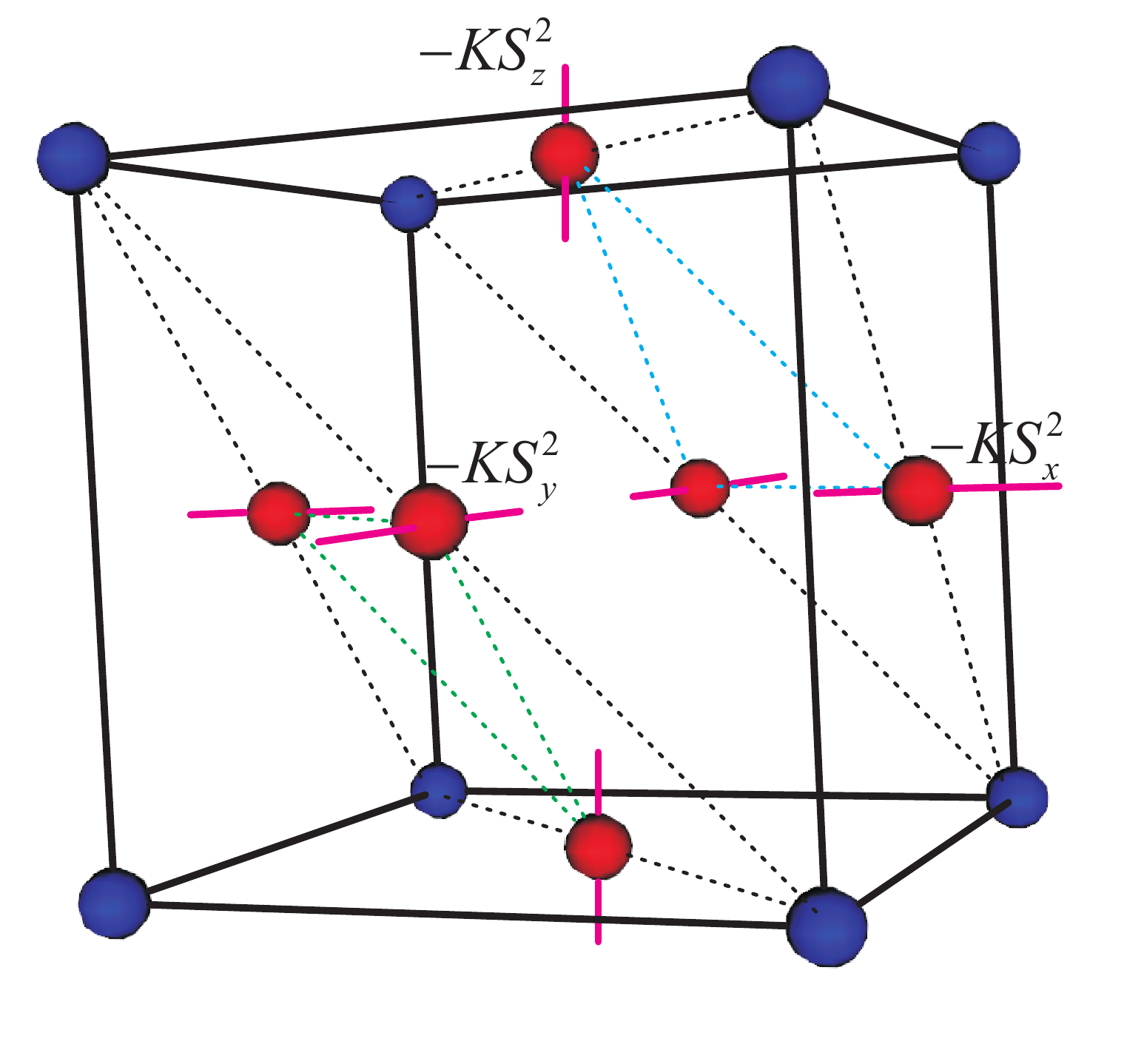}
\caption{Fcc atomic structure of the bulk IrMn$_3$. Ir ions (blue) at the vertices of the cube are not magnetic. Mn ions (red) form magnetic kagome lattices in $\{111\}$ planes. Local anisotropy axes are depicted.}
\label{fig:FccStructure}
\end{center}
\end{figure}

Recent studies of a truly 3D kagome structure embedded in a fcc lattice composed of ABC stacked kagome layers along $\langle 111 \rangle$ directions with eight AF NNs (depicted in Fig.~\ref{fig:FccStructure}) have been motivated by the ordered L1$_2$  phase of IrMn$_3$ and its sister compounds.   Neutron diffraction measurements \cite{tomeno1999} revealed the so-called T1 structure, in which Mn spins lie in $\lbrace 111 \rbrace$ planes along $\langle 112 \rangle$ directions. These correspond to the q=0 kagome spin configuration \cite{harris1992}, which may be characterized as three interpenetrating ferromagnetic sublattices. First principles calculations \cite{szunyogh2009} show an unusually strong effective cubic anisotropy term with local axes along $\langle 100 \rangle$ directions (see Fig.~\ref{fig:FccStructure}). In the same article, simulations using stochastic micromagnetic equations were reported along with confirmation of the T1 ordered state and high transition temperature $T_N\approx 1145$ K, consistent with experiment \cite{tomeno1999}.   Monte Carlo (MC) simulations of bulk IrMn$_3$ without anisotropy were conducted in Ref.~\cite{hemmati2012}, and with  cubic anisotropy added in Ref. ~\cite{leblanc2013} based on the  Hamiltonian obtained in Ref.~\cite{szunyogh2009}. They also confirmed the T1 state and showed that including cubic anisotropy changes the order of the  transition from first to second and gives rise to a net magnetization perpendicular to the kagome planes. The lifting of some of the degeneracies of the basic 2D structure when imbedded in the 3D fcc system, together with the effects of the cubic anisotropy, were shown to impact the spin wave dispersion curves \cite{leblanc2014}.

In contrast with thin ferromagnetic films \cite{jensen2006,bland2008}, layered structures composed of stacked frustrated AF lattices have received relatively little attention. Surface anisotropies (typically of the axial type normal to the film plane) can be substantial in magnitude and compete with those in the bulk \cite{popov2008,binder1992}. In the case of ferromagnetic films, it is well known that that the combined impact of finite-size effects associated with the reduced geometry as well as a reduction of coordination number and symmetry at the surface, can lead to significant changes in the spin structure relative to the bulk case. Most of the focus on AFs has involved helimagnets stabilized by either competing exchange interactions \cite{diep2015} or Dzyaloshinskii-Moriya interactions \cite{haraldsen2010,wilson2014}. Studies of surface effects in geometrically frustrated AFs have been isolated but reveal that profound changes can occur in the magnetic order \cite{langridge2014}, especially when sandwiched with a ferromagnet \cite{wang2013}. 
 
Alloys of Ir$_x$Mn$_{100-x}$, in both ordered and disordered phases, have found widespread utility in magnetic recording technology as the pinning AF layer, giving rise to the exchange bias (EB) phenomenon, in multi-layer spin valves \cite{tsunoda2009,tsunoda2010}. A neutron scattering investigation of 200 nm films of chemically ordered IrMn$_3$ suggests that the bulk T1 structure is preserved in this case but that chemically disordered films revealed a new magnetic structure characterized by a tilting of the moments away from cube face diagonals by about $45^\circ$ \cite{kohn2013}. This work also reports on studies of IrMn${}_3$/Fe(bcc) bilayers and exchange bias was measured. Of particular relevance to the present work is an electronic structure study of IrMn$_3$/Co(fcc) interfaces giving rise to a moderate surface anisotropy where the surface spin structure impacts the magnetic order in the first two or three layers and exhibits perpendicular coupling between Mn and Co spins \cite{szunyogh2011}.  This model, which includes Dzyaloshinskii-Moriya type antisymmetric exchange, served as the basis for a subsequent stochastic micromagnetic study of IrMn$_3$/Co(fcc) films that yielded EB, detected from calculated $MH$ hysteresis loops for a magnetic field applied perpendicular to the film plane \cite{yanes2013}.  We note that in spin-valve technology, EB is measured with the field applied in the film plane and the microscopic mechanism may be quite different.     
 

\begin{figure}[ht!]
\begin{center}
\includegraphics[width=0.7\textwidth]{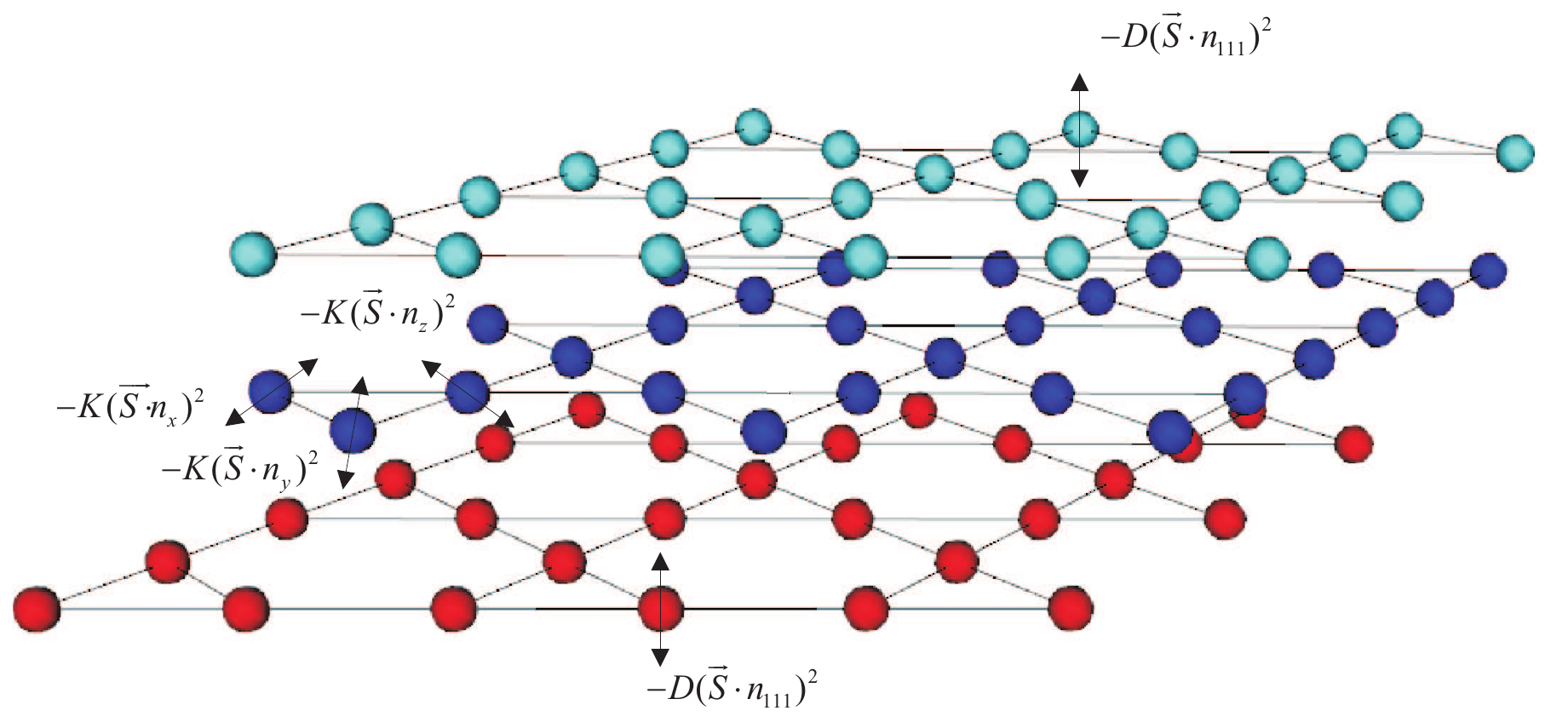}
\caption{Schematic model of the thin film with three layers.}
\label{fig:StackedKagome}
\end{center}
\end{figure}

In the present work, spin structures in the ground state as well as at finite temperature are examined using an effective field method and Metropolis MC simulations for three and six-layer ABC stacked kagome layers forming (111) planes. Interior layers are assigned a cubic anisotropy (as in the bulk case) whereas surface layers are given a uniaxial aniostropy $-DS^2_{z'}$ (where $\bf{\hat z'}$ is perpendicular to the (111) plane), as illustrated in Fig.~\ref{fig:StackedKagome}. The magnetic order of the surface and interior layers, as well as the N\'eel temperature, are studied as a function of $D$. The relative impact of the two types of anisotropy on the reduction of the fundmental kagome spin degeneracy is discussed. Special attention is given to the magnetization with a view of its potential relevance to the EB effect.     


The remainder of the paper is organised as follows. In Sec. 2, a model for thin films of IrMn${}_{3}$  is described. In Sec. 3, the ground state of the system is analyzed analytically with results supported by calculations using an effective field method as well as MC simulations (with details provided in an appendix). In Sec. 4, MC simulation results on the specific heat are presented for the thin films at surface and interior layers. Corresponding results for the order parameters and magnetization are shown in Sec. 5 and Sec. 6, respectively. Our conclusions are given in Sec. 7.

\section{The Model}

We consider thin films composed of ABC stacked kagome layers. Inspired by the work in Ref. ~\cite{szunyogh2009} and taking into account effects of axial surface anisotropy, we consider the Hamiltonian 
\begin{eqnarray}
\label{hamiltonian}
\mathcal{H} = &- J\sum_{\langle \mathrm{NN} \rangle} \vec{S_i} \cdot \vec{S_j} - K\sum_{\gamma\in\{x,y,z\}} \sum_{i\in\mathrm{bulk}} (\vec{n}_\gamma \cdot \vec{S}_i)^2  \nonumber \\
 &-D \sum_{i\in\mathrm{surface}}(\vec{n}_{111}\cdot \vec{S}_i)^2,  
\end{eqnarray}
where  $\vec{n}_\gamma$ are unit vectors along $x$, $y$ and $z$ axes of the conventional fcc cell of the corresponding bulk lattice and  $\vec{n}_{111}$ is a unit vector normal to the surface  (in the [111] direction). Here, the isotropic exchange interaction ($J$) is restricted to the NN interactions. It is also assumed that the exchange interaction is the same for six NN surface spins and the eight NN interior spins. The second term includes the contribution of the effective on-site cubic anisotropy \cite{szunyogh2009}, and the third term represents the assumed axial surface anisotropy. For the remainder of this paper, dimensionless units are defined by taking $J=-1$. In the particular case of IrMn$_3$, the cubic anisotropy constant has been estimated as 10\% of value of $|J|$ \cite{szunyogh2009} so that we assume here for convenience the value  $K=0.1$ throughout. The surface anisotropy constant $D$ is varied over a range of positive values thus defining an easy axis which is perpendicular to the plane, which is common in thin films \cite{jensen2006,bland2008}.   Note that a negative value of $D$ would not influence the spin structure in the ground state except to force the surface spins to lie on the $(111)$ plane.

\section{Ground State}
\label{GroundState}
The following analysis of the ground state spin structures corresponding to the model described above is carried out for three and six layer films in the Cartesian coordinate system with axes coinciding with the conventional unit cell of the corresponding infinite fcc kagome lattice. While the generalization of the analysis of ground state properties to a larger number of layers is straightforward there is little qualitative change in the spin structure and the extension to the general case is only discussed briefly towards the end of this section.

Based on the results from low temperature MC simulations we assume here, as in the bulk case \cite{leblanc2013}, that there are only three distinct spin directions in each layer. Denoting the surface spins by $\mathbf{S}_1, \mathbf{S}_2, \mathbf{S}_3$,  and the interior spins by $\mathbf{M}_1, \mathbf{M}_2, \mathbf{M}_3$ such that spins with the index $1$ have as NNs only spins with indices $2$ and $3$, spins with the index $2$ has NNs only those with indices $1$ and $3$, etc. The local cubic anisotropy axes for the interior spins $M_1$, $M_2$, and $M_3$ are $\hat{x}$, $\hat{y}$, and $\hat{z}$, respectively. For the three layer case the expression for the energy may then be written as 
\begin{eqnarray} 
E= &- \frac{4}{9} J \left(\mathbf{S}_1 \cdot \mathbf{S}_2+\mathbf{S}_1 \cdot \mathbf{S}_3+\mathbf{S}_2 \cdot \mathbf{S}_3\right) \nonumber \\
&- \frac{2}{9} J \left(\mathbf{M}_1 \cdot \mathbf{M}_2+\mathbf{M}_1 \cdot \mathbf{M}_3+\mathbf{M}_2 \cdot \mathbf{M}_3\right) \nonumber  \\
&- \frac{2}{9} J \left[\mathbf{M}_1 \cdot ( \mathbf{S}_2+\mathbf{S}_3) + \mathbf{M}_2 \cdot (\mathbf{S}_1+\mathbf{S}_3) + \mathbf{M}_3 \cdot (\mathbf{S}_1+\mathbf{S}_2)  \right] \nonumber \\
&-\frac{2}{27} D [(S_{1x}+S_{1y}+S_{1z})^2 + (S_{2x}+S_{2y}+S_{2z})^2 \nonumber \\
& +(S_{3x}+S_{3y}+S_{3z})^2 ] -\frac{1}{9} K \left( {M_{1x}}^2+{M_{2y}}^2+{M_{3z}}^2  \right).
\label{eqn:energy_per_spin}
\end{eqnarray}
To study minimum energy spin configurations based on Eq. \ref{eqn:energy_per_spin} for $K=0.1$ as a function of $D$ we first consider the two case limiting cases $D \ll K \ll |J|$ and  $K \ll |J| \ll D$ as these limits can be treated analytically. Numerical results based on an effective field method \cite{walker1980} are also presented over a range of values of $0< D \le 10$ that interpolate between these two limiting cases. 

\begin{figure}[!htbp]
		\includegraphics[width=\linewidth]{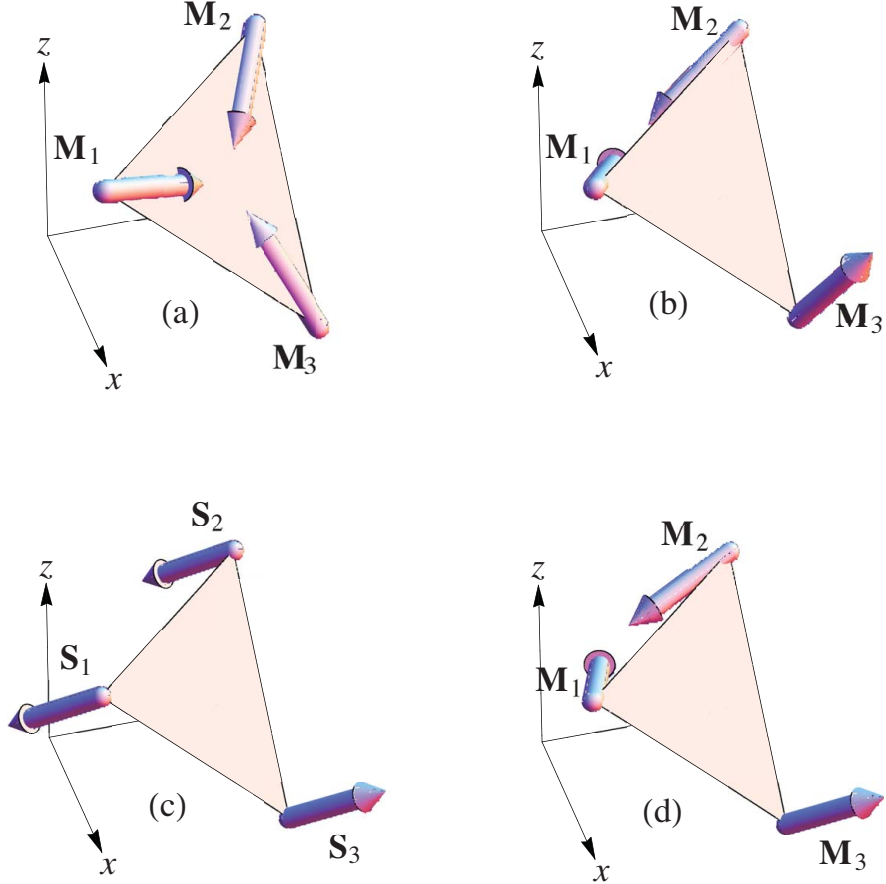}
\caption{ Ground state spin configurations for various limiting cases of the parameters $D$, $K$ and $J$ are shown schematically. The triangular surface shown in all the above figures indicates the $(111)$ surface. 
(a) The ground state spin configuration for $D=K=0$ refered to as $\Phi_0$ in the text. 
(b) The ground state spin configuration for both the interior spins ($\mathbf{M}_i$)  and the surface spins ($\mathbf{S}_i$) for the case $D=0^{+}$, $K = 0.1$ and $|J|\gg K$,
(c) ground state spin configuration for the surface spins ($\mathbf{S}_i$) for the case $D \gg |J|$.  
(d) The ground state spin configuration for the interior spins ($\mathbf{M}_i$) for the case $K=0.1$ and $K \ll |J| \ll D$.}
\label{fig:spinConfigs}
\end{figure}

Consider first the limit $D\ll K\ll J$ and begin by with the specific case $D=K=0$.  In the absence of anisotropy the spins will align to minimise the exchange energy such that $\mathbf{S}_i= \mathbf{M}_i $ and 
\begin{eqnarray}
\mathbf{S}_i\cdot\mathbf{S}_j = \mathbf{M}_i\cdot\mathbf{S}_j = \mathbf{M}_i\cdot\mathbf{M}_j = -\frac12\quad \text{for}\quad i\ne j \label{constraint}
\end{eqnarray}
with $E_\mathrm{ex} = -5J/3$. We note that the minimum exchange energy is highly degenerate since it is invariant under any global rotation of the spins.  To understand this degeneracy we define ground state $\Phi_0$ in which the spins all lie in the $(111)$ plane with 
\begin{subequations}\label{E:gp1}
\begin{align}
\mathbf{S}_1=\mathbf{M}_1=\frac{1}{\sqrt{6}}\left(-1,2,-1\right)  \\
\mathbf{S}_2=\mathbf{M}_2=\frac{1}{\sqrt{6}} \left(2,-1,-1\right) \\
\mathbf{S}_3=\mathbf{M}_3=\frac{1}{\sqrt{6}}\left(-1,-1,2\right)
\end{align}
\end{subequations}
This $120^0$ spin configuration is shown schematically in Fig.~\ref{fig:spinConfigs}(a). From this we can construct a sequence of ground state spin configurations by a rotation of $\Phi_0$ around the axis $\mathbf{u} =(1,-1,0)/\sqrt{3}$ by some angle $\theta$ to generate a new ground state $\Phi(\theta)=(\mathbf{M}_i(\theta),\mathbf{S}_i(\theta))$ where $\mathbf{M}_i(\theta)$ and $\mathbf{S}_i(\theta)$ are defined by 
\begin{eqnarray}
 \mathbf{S}_i(\theta)=\mathbf{M}_i(\theta) =\mathbf{R}_\mathbf{u}(\theta)\cdot \mathbf{M}_i
\label{thetaPhi}
\end{eqnarray} 
with $\mathbf{R}_\mathbf{u}(\theta)$ representing the rotation tensor.
Figures~\ref{fig:spinConfigs}(b) and ~\ref{fig:spinConfigs}(c) show the states $\Phi(70.582^\circ)$ and $\Phi(90^\circ)$, respectively, discussed below.

This degeneracy is broken by both the cubic anisotropy of the interior spins and the axial anisotropy of the surface spins. We consider first the case of finite $K=0.1$ and  $D=0$. The normalised cubic anisotropy energy $\epsilon_\mathrm{cubic}(\theta) = -(M^2_{1x}(\theta) + M^2_{2y}(\theta) + M^2_{3z}(\theta))$ is plotted in Fig. 4 over the range $-\pi \le \theta \le \pi$. From Fig. 4 we see that  $ \epsilon_\mathrm{cubic}(\theta)$ is two fold degenerate with minima corresponding to $\theta_1\approx 70.6^\circ$ and $\theta_2 = 180^\circ$ with $\epsilon_\mathrm{cubic}(\theta_1)=\epsilon_\mathrm{cubic}(\theta_2)=-2$. We note that the spin configuration corresponding to the state $\Phi(\theta_1)$ lie in the $(11\bar1)$ plane, while for the $\Phi(\theta_2)$ they lie in the $(111)$. 
The specific spin configuration for the $\Phi(\theta_1)$ state is given by
\begin{subequations}\label{E:gp2}
\begin{align}
\mathbf{S}_1&=\mathbf{M}_1=\frac{1}{\sqrt{6}}\left(-2,1,-1\right)  \\
\mathbf{S}_2&=\mathbf{M}_2=\frac{1}{\sqrt{6}} \left(1,-2,-1\right) \\
\mathbf{S}_3&=\mathbf{M}_3=\frac{1}{\sqrt{6}}\left(1,1,2\right)
\end{align}
\end{subequations}
and is shown schematically in Fig.~\ref{fig:spinConfigs}(b). Thus the effect of the the cubic anisotropy is to reduce the continuous degeneracy of a global rotation to a eightfold degeneracy (as we could have equivalently used the [011] or [101] as our axis of rotation, or reversed the spins ($\mathbf{S}_i \to -\mathbf{S}_i$ in Eqs.~\ref{E:gp1}).  

\begin{figure}[!htbp]
		\includegraphics[width=0.75\linewidth]{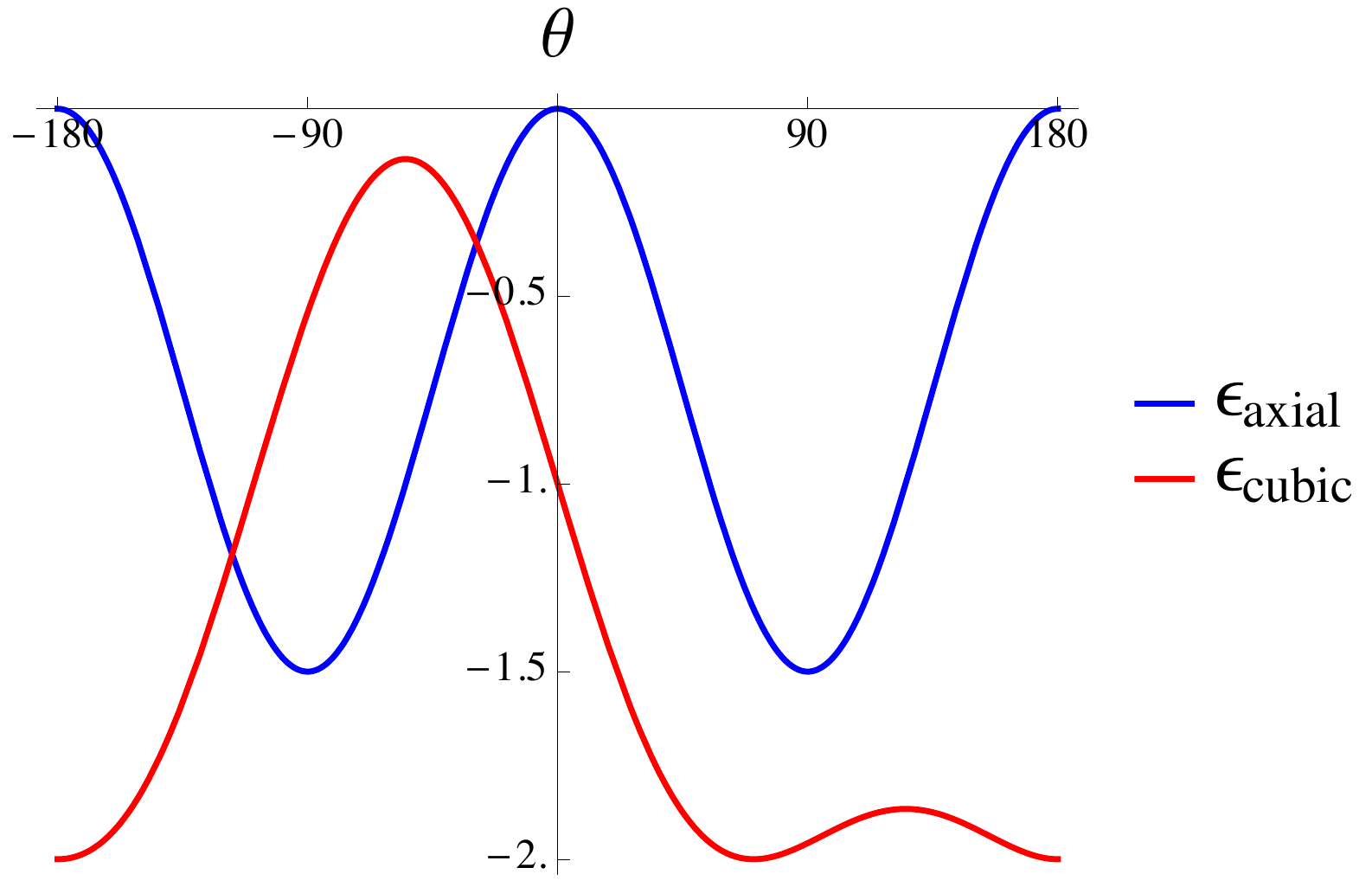}	
\caption{Plot of the nomalised anisotropy energies $\epsilon_\mathrm{cubic}=-\sum_{i}(M^2_{1x}(\theta) + M^2_{2y}(\theta) + M^2_{3z}(\theta))$ and $\epsilon_\mathrm{axial}=-\sum_i(\mathbf{S}_{i}(\theta).\hat{\mathbf{n}}_{111})^2 $ calculated for the states $\Phi(\theta)$ as a function of the rotation angle $\theta$.}
\label{Angles}
\end{figure}

A plot of the normalised axial anisotropy energy $\epsilon_\mathrm{axial}(\theta) = -((\mathbf{S}_{1}.\hat{\mathbf{n}}_{111})^2 +(\mathbf{S}_{2}.\hat{\mathbf{n}}_{111})^2 + (\mathbf{S}_{3}.\hat{\mathbf{n}}_{111})^2)$ as a function of $\theta$ is also presented in Fig~\ref{Angles}. The graph shows two minima at $\theta=\pm90^\circ$ in which the spins lie in the $(11\bar{2})$ plane with $\mathbf{S}_3$ directed along $\hat{\mathbf{n}}_{111}$ axis, as shown in Fig. 3(c), and two maxima at $\theta =0$ and $180^\circ$ in which the spins lie in the $(111)$ plane. The two curves show that the surface axial anisotropy will break the degeneracy of the $\Phi(\theta_1)$ and the $\Phi(\theta_2)$ states; $D = 0^{+}$ will select $\Phi(\theta_1)$ as the ground state while $D = 0^{-}$ will select the $\Phi(\theta_2)$ state. Since we are interested only in the case $D>0$, $\Phi(\theta_1)$ is the relevant ground state when $D$ is small yet finite. 

The above results apply when $D = 0^{+}$, $K =0.1$ for $K \ll J$. In order to compare the ground state spin configuration obtained in these limiting cases with that obtained for small but finite $D$ and for $K = 0.1$,  we have calculated the ground state to second order in $K/J$ and $D/J$. The details of the perturbation theory are described in Appendix A and the results tabulated in Table~\ref{NNangles}.  The angles between NN spins calculated for $K=0.1$ and $D=0.02$ are very close to, but not exactly equal to, $120^\circ$. However, the angle between the spins and the surface normal is significantly different (eg. in the limit $D\to 0^{+}$ we have, with an overline denoting the angle between two vectors,  $\widehat{\hat{\mathbf{n}}_{111}\mathbf{S}_3}= 19.5^\circ$ and $\widehat{\hat{\mathbf{n}}_{111}\mathbf{S}_1}= 118.1^\circ$) and clearly shows the effect of the surface axial anisotropy tilting the spins towards the $\hat{\mathbf{n}}_{111}$ axis while essentially maintaining the characteristic  $120^\circ$ kagome spin structure between the NNs. 

\begin{table}
\begin{tabular}{|c|c|c|c|c|c|c||c|}
\hline
&$\mathbf{S}_1$ & $\mathbf{S}_2$ & $\mathbf{S}_ 3$& $\mathbf{M}_1$ & $\mathbf{M}_2$ & $\mathbf{M}_ 3$&$\hat{\mathbf{n}}_{111}$\\\hline
$\mathbf{S}_1$ &&$119.55^\circ$&$120.23^\circ$&&$119.64^\circ$&$120.23^\circ$&$119.9^\circ$\\\hline
$\mathbf{S}_2$ &$119.55^\circ$&&$119.64^\circ$&$119.64^\circ$&&$120.23^\circ$&$119.9^\circ$\\\hline
$\mathbf{S}_3$ &$120.23^\circ$&$120.23^\circ$&&$120.13^\circ$&$120.13^\circ$&&$7.8^\circ$\\\hline
$\mathbf{M}_1$ &&$119.64^\circ$&$120.13^\circ$&&$119.72^\circ$&$120.11^\circ$&$119.9^\circ$\\\hline
$\mathbf{M}_2$ &$119.64^\circ$&&$120.13^\circ$&$119.72^\circ$&&$120.11^\circ$&$119.9^\circ$\\\hline
$\mathbf{M}_3$ &$120.23^\circ$&$120.23^\circ$&&$120.11^\circ$&$120.11^\circ$&&$9.1^\circ$\\\hline
\end{tabular}
\vspace{0.5cm}
\caption{Angles between NN spins and the surface normal, $\hat{\mathbf{n}}_{111}$, in ground state configuration calculated for $K=0.1$ and $D=0.02$  using perturbation method desrcibed in Appendix A.} \label{NNangles}
\end{table}

The other limit that admits an analytical solution is the case $K \ll J \ll D$. Here, surface anisotropy will dominate and  hence the surface spin structure will be of the Ising type with the spins aligned along the $\hat{\mathbf{n}}_{111}$ axis, with $\mathbf{S}_3$ pointing away from the plane and $\mathbf{S}_1$ and $\mathbf{S}_2$ pointing towards the interior, as shown in Fig.~\ref{fig:spinConfigs}(c). Whereas in the previous case the dominant nature of the exchange required that $\mathbf{S}_i = \mathbf{M}_i$, such an assumption cannot be made here. Instead, noting that the interlayer exchange only couples $\mathbf{S}_2$ and $\mathbf{S}_3$ to $\mathbf{M}_1$, and given that $\mathbf{S}_2=-\mathbf{S}_3$, then the net exchange field acting on $\mathbf{M}_1$ will be zero. A similar argument may be applied to $\mathbf{M}_2$. On the other hand $\mathbf{M}_3$ will experience a net exchange field from $\mathbf{S}_1$ and $\mathbf{S}_2$ which is finite.  Based on this argument,  the expression for the energy given by Eq.~(\ref{eqn:energy_per_spin}) reduces to
\begin{eqnarray}
E = & -\frac{2}{9} J (\mathbf{M}_1 \cdot \mathbf{M}_2+\mathbf{M}_1 \cdot \mathbf{M}_3+\mathbf{M}_2 \cdot \mathbf{M}_3)- \frac{2}{3} D +\frac{4}{9}J  \nonumber \\
&- \frac{1}{9} K \left( {M_{1x}}^2+{M_{2y}}^2+{M_{3z}}^2  \right)  + \frac{4}{9\sqrt{3}} J  (M_{3x}+M_{3y}+M_{3z}).
\label{eqn:Energy per spin big D}
\end{eqnarray}
From here it immediately follows that for $K \ll J$ the ground state solution for the middle layer spins will be of the form given by Eq.~\ref{thetaPhi} with $\mathbf{M_3}$ pointing in the $[111]$ direction as shown schematically in Fig.~\ref{fig:spinConfigs}(d).

The above analysis shows that in the limit $K \ll J$, the effect of the surface axial anisotropy is to transform the essentially antiferromagnetic $q=0$ ground state with the spins lying in the $(11\bar{1})$ plane in the limit $D\to 0^{+}$, into a state in which the surface spins are aligned perpendicular to the (111) plane in a collinear ferrimagnetic configuration ($\mathbf{S}_3$ up, $\mathbf{S}_1$ and $\mathbf{S}_2$ down), while the interior spins are simply rotated by $90^\circ$ around the $(1\bar{1}0)$ axis so that  $\mathbf{M}_1$, $\mathbf{M}_2$ and $\mathbf{M}_3$ are aligned parallel to the $(11\bar{2})$ and $\mathbf{M}_3$ is aligned parallel to $\mathbf{S}_3$ (such that 
the angle betwen NN interior spins is $120^\circ$ in the limit $D \to \infty$).

\begin{figure}[!htbp]
		\centering
		\vspace{0.5mm}
		\includegraphics[width=0.5\linewidth]{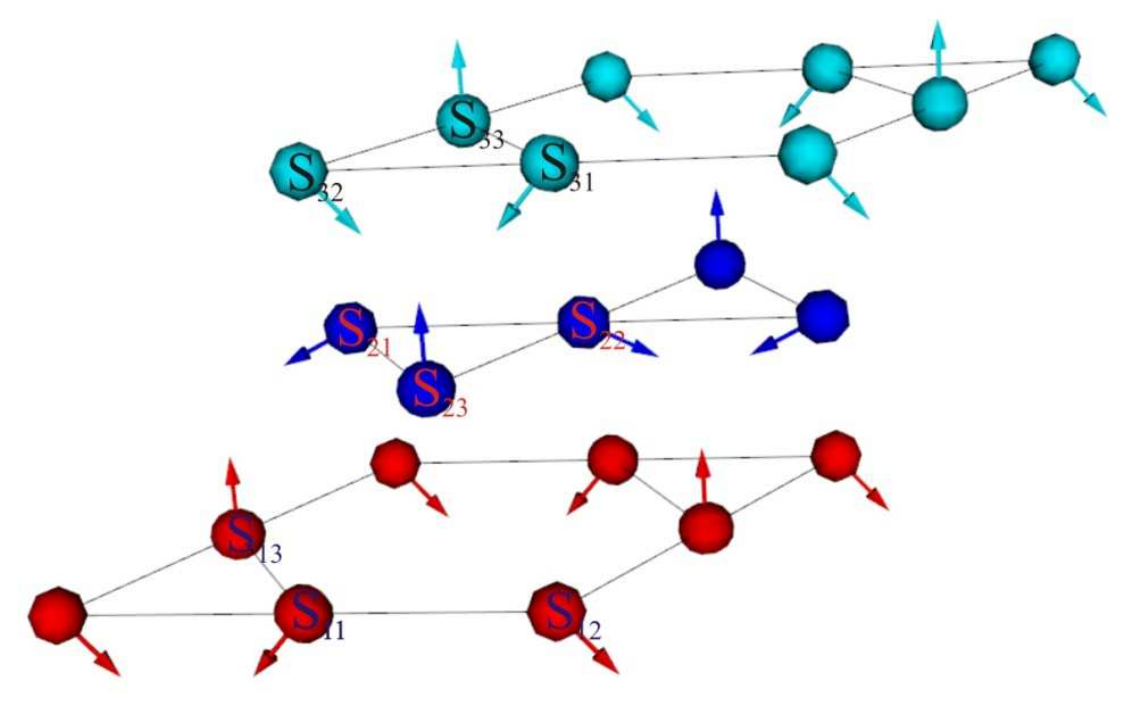}
\caption{Ground state of the 3-layer system with $D=1$ and $K=0.1$.}
\label{fig:TypeOfIons}
\end{figure}

To determine how the system transforms between the two limiting cases discussed above we have computed ground state of Eq. \ref{eqn:energy_per_spin} using the effective field method of Ref. \cite{walker1980} for $K=0.1$ and $0 < D < 10$. Calculations for both $L = 3$ and $6$ were performed on lattices of size  $6 \times 6 \times L$ with periodic boundary conditions in the lateral direction. A schematic of the ground state configuration for $K=0.1$ and $D=1.0$ is shown in Fig. \ref{fig:TypeOfIons} for the three layer case. 

\begin{figure}[!htbp]
		\centering
		\includegraphics[width=0.7\linewidth]{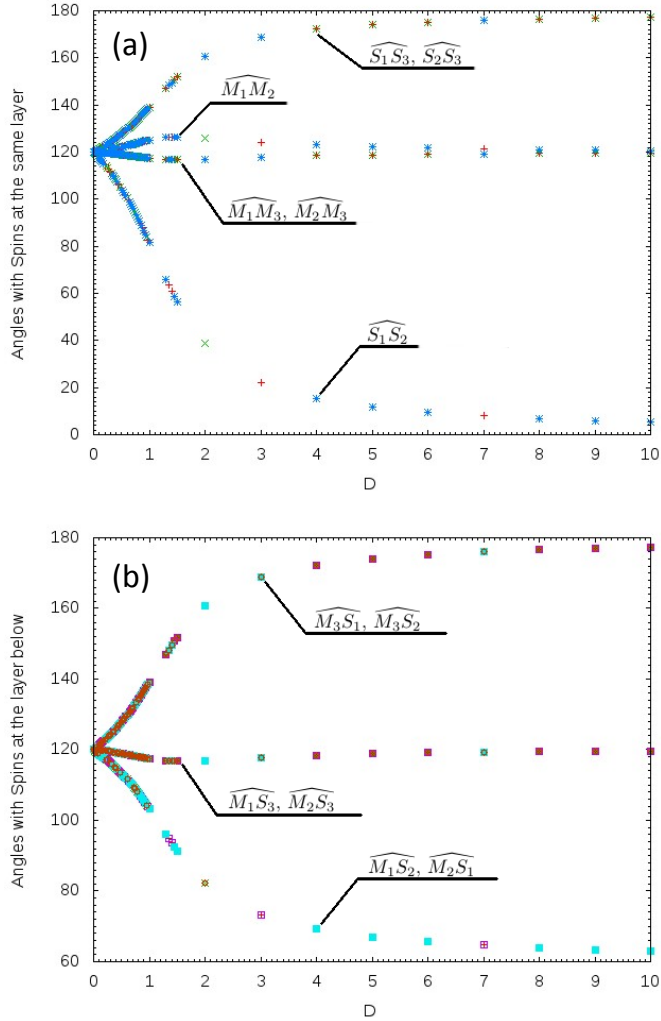}
\caption{Ground state angles for the three layer system (a) between spins in the same layer and (b) between spins in adjacent layers plotted as a function of $D$ for $K=0.1$.}
\label{fig:Angles}
\end{figure}

To illustrate the dependence of the ground state spin configuration on $D$,  the angles between the NN spins within the same layer, denoted by $\widehat{S_i S_j}$ and $\widehat{M_i M_j}$, and adjacent layers, denoted by $\widehat{M_i S_j}$, are plotted as a function $D$ for the three layer case in Fig. \ref{fig:Angles}. The angles were calculated for each pair of adjacent spins and then averaged over the lattice.  All of these ground-state results were verified by zero-T MC simulations using lattices $18 \times 18 \times 3$ at a temperature $T = 10^{-6}$. 

\begin{figure}[!htbp]
		\centering
		\includegraphics[width=0.6\linewidth]{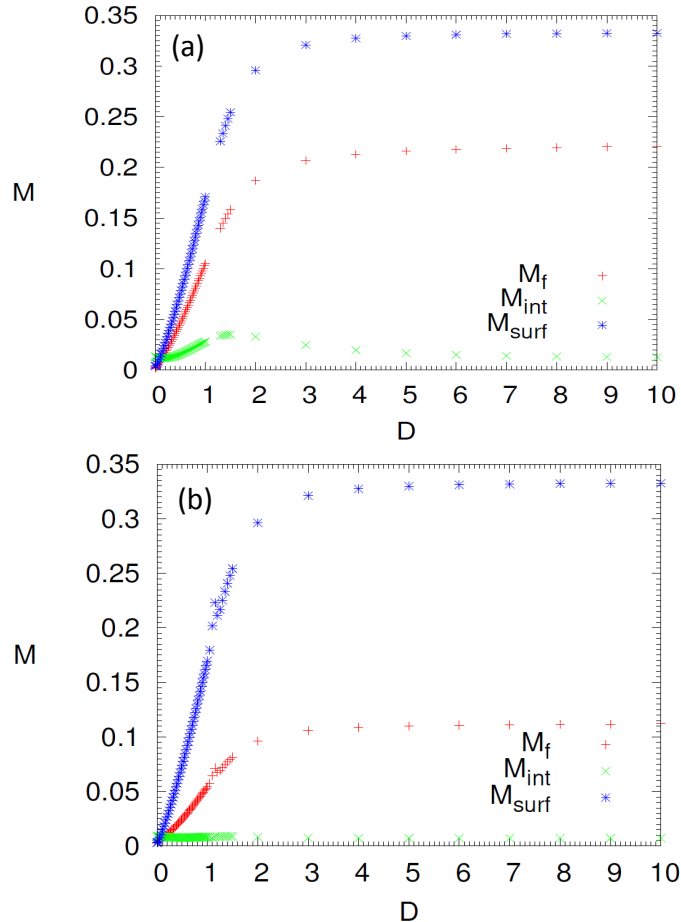}
\caption{A plot of the magnetization per spin of the surface and interior spins as a function of $D$ for both (a)  three layer and (b) six layer cases.}
\label{fig:GSmagnetization}
\end{figure}

The significance of these results from the perspective of exchange bias is clearly demonstrated in Fig.~\ref{fig:GSmagnetization} in which the ground state surface magnetization per unit spin together with magnetization per unit spin of the interior layer(s) for both $L=3$ and $6$ are plotted.  The data show a significant surface magnetization which increases with increasing $D$, saturating at $M_\mathrm{surf} = 1/3$ per spin  for both the three and six layer cases. In the three layer case,  the magnetisation of the interior layer initially increases from effectively zero to a maximum value  of $0.04$ at $D \approx 1.4$, decreasing monotonically thereafter. For the six layer case,  the interior magnetisation is effectively zero ($M_\mathrm{int} < 0.0009$) over the entire range of $D$ . Thus the effect of the axial surface anisotropy is  to induce a ferrimagnetic surface spin configuration with a finite surface magnetization enclosing an antiferromagnetic kagome spin configuration within the interior layer(s). 

\section{Monte Carlo Simulations}

The MC simulations presented in this section have been carried out for systems consisting of $L$ layers, with $L=3$ and $6$, of area $18\times18$. In all cases, the system is initialized to some random spin configuration and a simulation performed at some suitably high initial temperature (typically T=2.5). The system is then cooled by using the final state of the previous simulation as the initial configuration for the subsequent simulation, as the temperature is lowered.  Between $2.0 \times 10^5$ and  $2.5 \times 10^5$ MC steps (MCS) were used, with the initial 10\% being discarded for equilibration.  

\begin{figure}[!htbp] 
		\centering
		\includegraphics[width=0.7\linewidth]{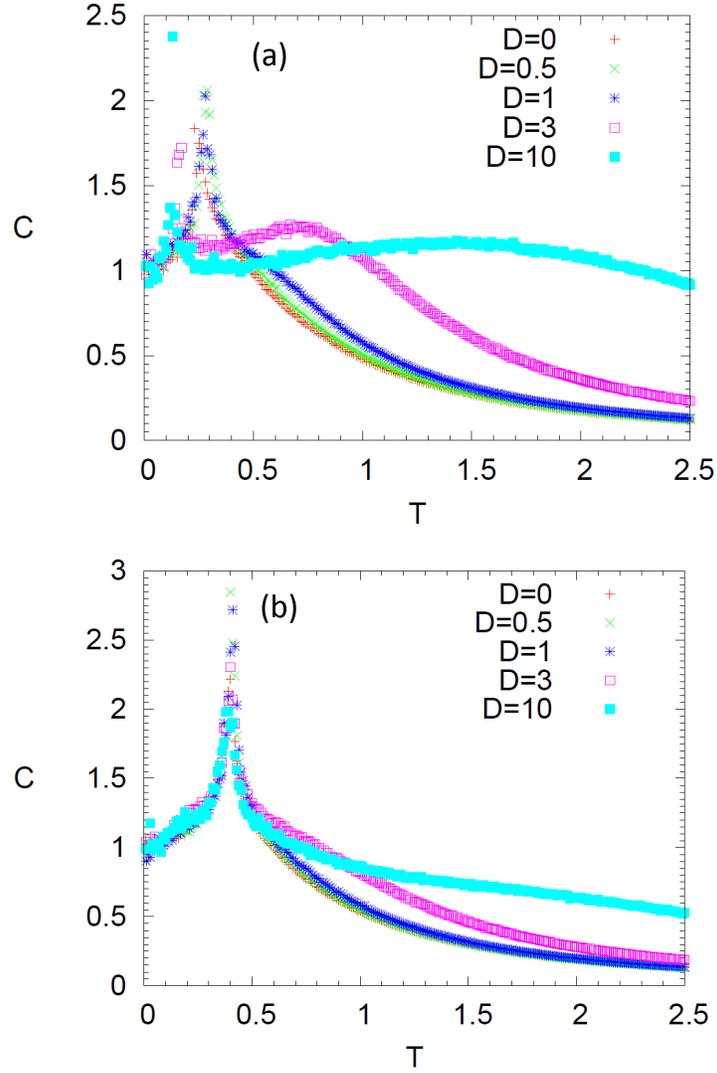}
\caption{Specific heat for (a) three layer and (b) six layer films. }
\label{fig:Cv}
\end{figure}

Figure~\ref{fig:Cv} shows the specific heat per spin for the three  and six layer systems with $K=0.1$, as a function of temperature and surface anisotropy $D$. The sharp peaks correspond to  the onset of magnetic order.   We note that  for sufficiently large $D$ the specific heat data for the three layer system exhibit a broad shoulder at high temperature that does not appear in the six layer case. 

\begin{figure}[!htbp] 
		\centering
		\includegraphics[width=0.7\linewidth]{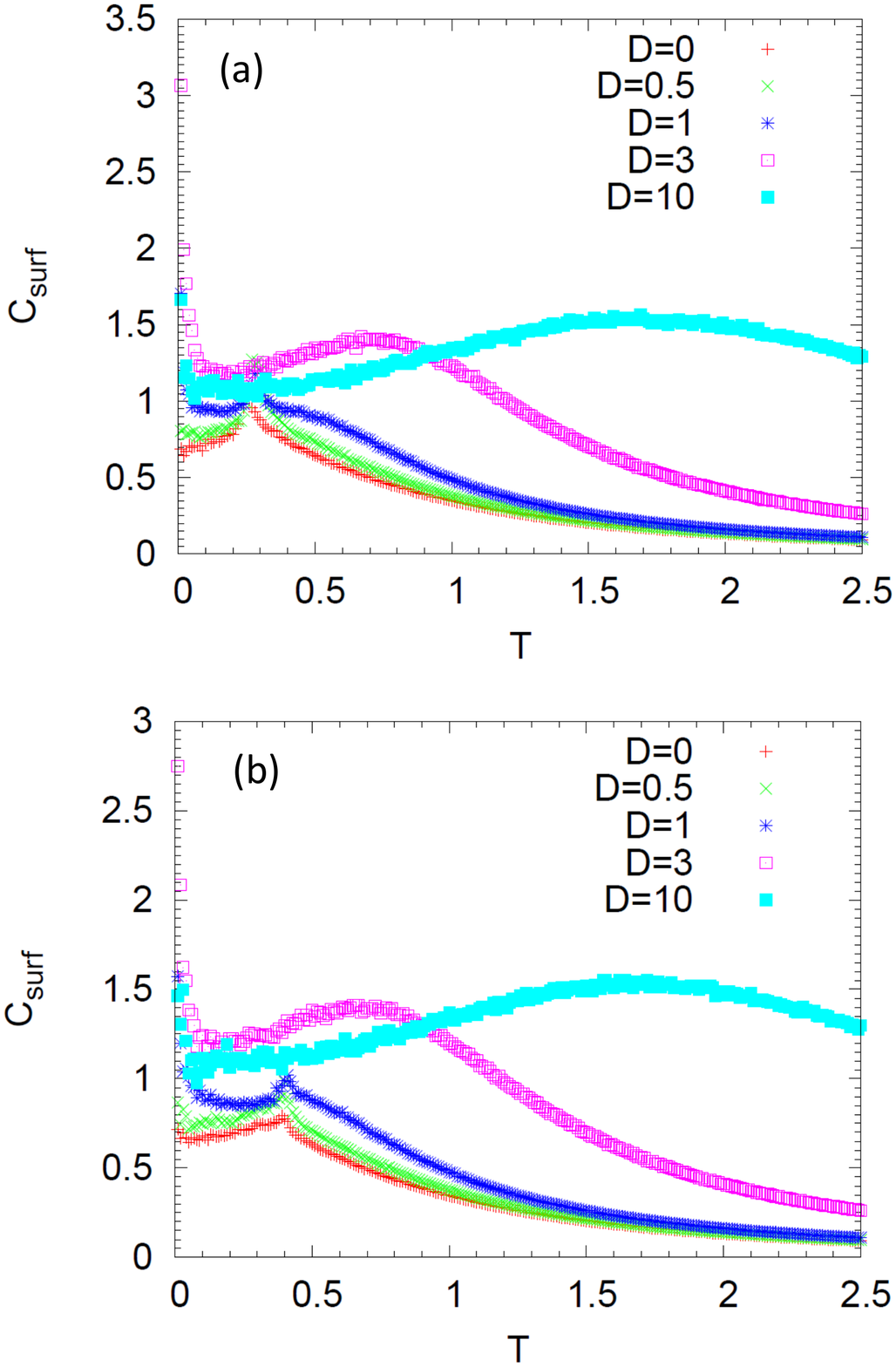}
\caption{Contribution to the specific heat from surface layers for (a) three layer films and (b) six layer films.}
\label{fig:Cvsurface}
\end{figure}

\begin{figure}[!htbp] 
		\centering
		\includegraphics[width=0.7\linewidth]{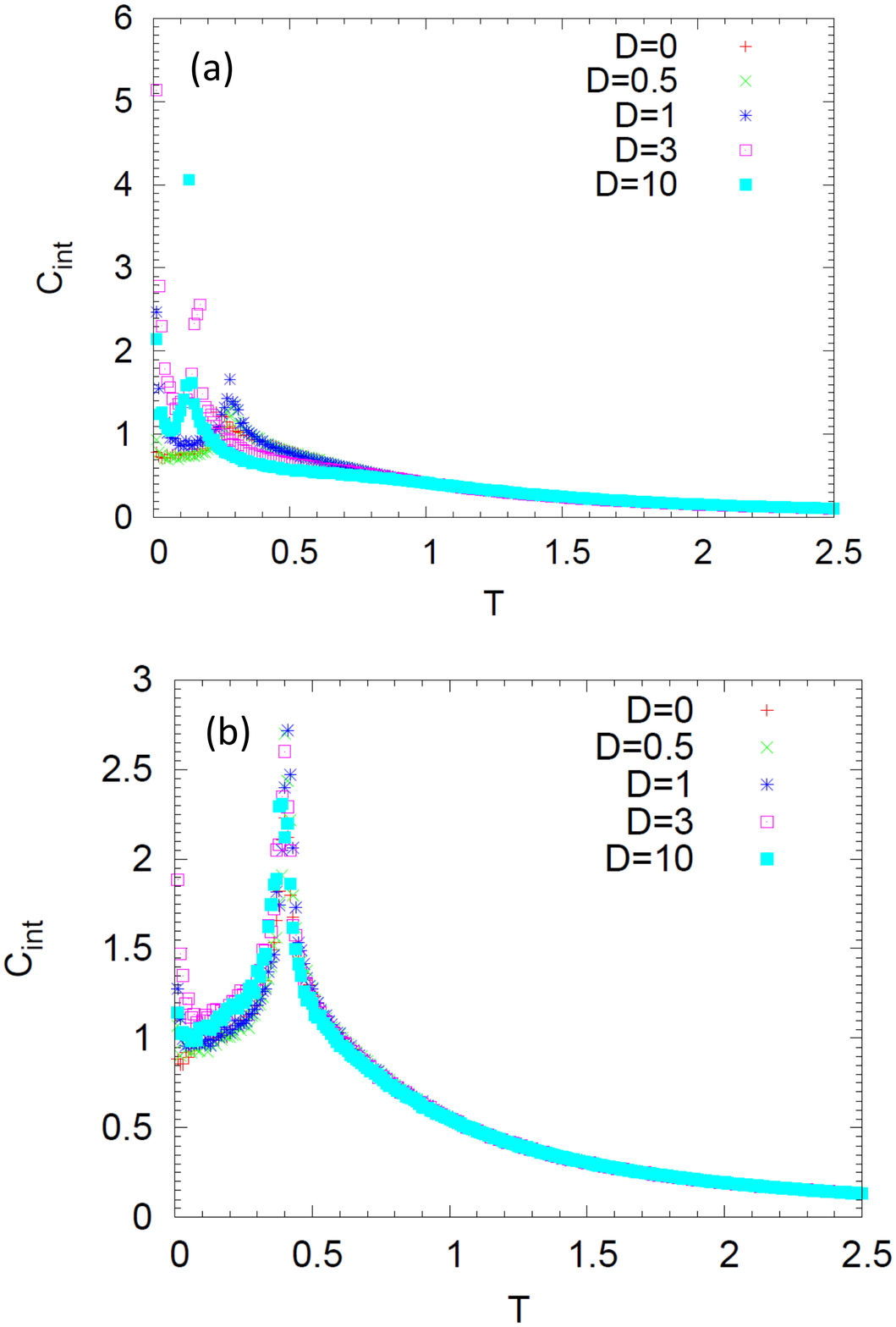}
\caption{Contribution to the specific heat from interior layers for (a) three layer films and (b)  six layer films.}
\label{fig:Cvinterior}
\end{figure}

That the high-$T$ shoulder is observed only for three-layer film suggests that it is a surface effect. To confirm this we introduce the quantities $C_\mathrm{surf}$ and $C_\mathrm{int}$ defined as the contributions to $C$ arising from the energy fluctuations of the surface and the interior spins, respectively, These are given by the following expressions
\begin{eqnarray}
C_\mathrm{surf} &= k_B\beta^2 \left ( \langle E_\mathrm{surf}^2 \rangle - \langle E_\mathrm{surf} \rangle^2 \right )\\ 
C_\mathrm{int} &= k_B\beta^2 \left ( \langle E_\mathrm{int}^2 \rangle - \langle E_\mathrm{int} \rangle^2 \right ),
\end{eqnarray}
where $E_\mathrm{surf}$ and $E_\mathrm{int}$ denote the energy associated with the surface and interior layers respectively.  The data for $C_\mathrm{surf}$ and $C_\mathrm{int}$ as a function of $T$ are shown for in Figs.~\ref{fig:Cvsurface} and ~\ref{fig:Cvinterior} and suggest that the high-T shoulder may be attributed to surface term in the Hamiltonian. This is analogous to the Schottky anomaly [28] observed in simulations performed at large $K$ values  in bulk IrMn3 \cite{leblanc2013}.

\begin{figure}[ht]
\begin{center}
\includegraphics[width=0.6\linewidth]{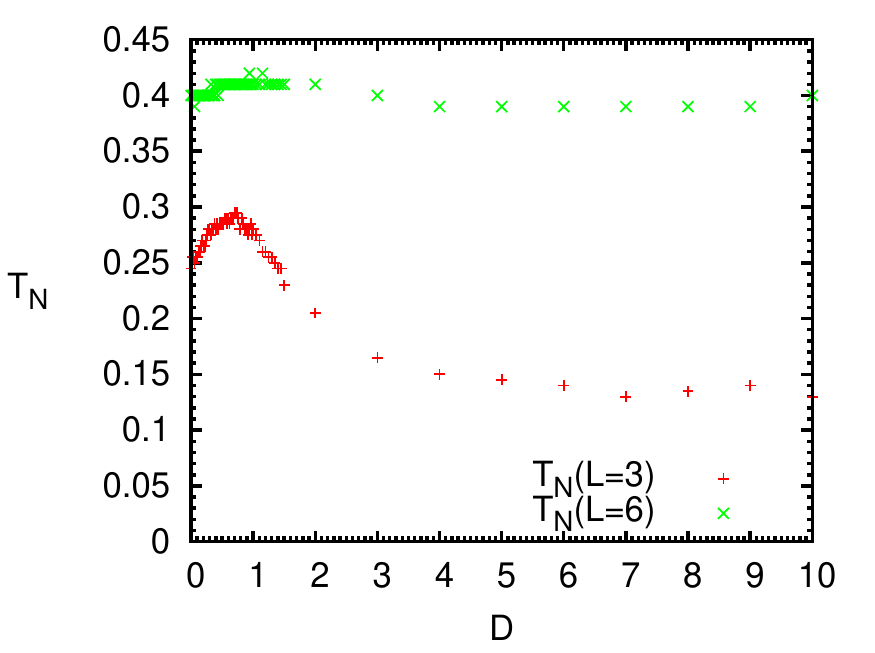}
\caption{Transition temperature deduced from specific heat plots.}
\label{fig:TransTvsDKIntC18_18_L}
\end{center}
\end{figure}

Figure~\ref{fig:TransTvsDKIntC18_18_L} shows the transition temperature for both three and six layer cases as a function of $D$ estimated from the peaks in the specific heat. As expected for thin-film systems \cite{binder1992}, the ordering temperatures estimated from the specific heat data are, in all cases, less than that of the bulk system, $T_N \simeq 0.52$ \cite{leblanc2013}. For the three layer film, there is well formed maximum in the specific heat near $D=0.7$, which also exists for the six layer case but is not so pronounced. We also note that as $D\rightarrow \infty$, $T_N$ does not tend to $0$, but to some finite value because the $D$ term acts only on the surface spins and the system can still establish order through coupling to the interior layers. By way of comparison, for $K=D=0.1$, the transition temperature is about $0.52$ in the 3D case.  The thin-film results above indicate that it has values of about $0.41$, for $L=6$ and $0.25$ for $L=3$. Comparing Figs. ~\ref{fig:Cv} and ~\ref{fig:Cvinterior} it is evident that the interior spins in the six layer case provide the main contribution to the magnetic order, as expected.  

\section{Order Parameter}

\begin{figure}[!htbp] 
		\centering
		\includegraphics[width=0.7\linewidth]{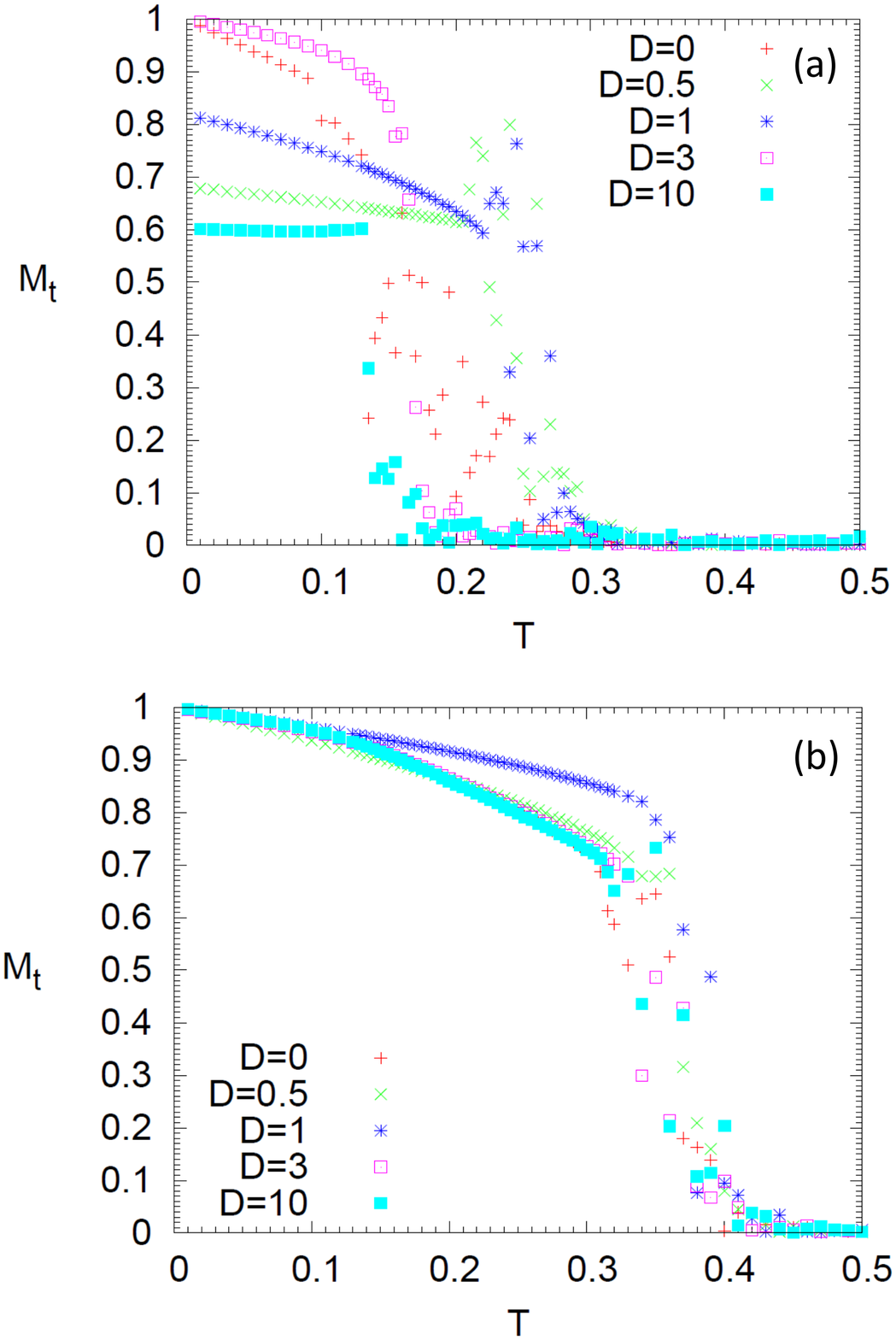}		
\caption{Order parameter $M_{t}$ of the  films with (a) three layers  and (b) six layers obtained from the cooling cycle.}
\label{fig:Mt}
\end{figure}

The ground states of the form given by Eqs. \ref{E:gp1} consist of three interpenetrating ferromagnetic lattices and therefore the onset of order can be characterised by the order parameter \cite{leblanc2013} 
\begin{eqnarray} 
M_t&=\frac{1}{N} \sum_{\mathrm{layers}} \sum_{\gamma} \left| \left\langle \sum_{i \epsilon \gamma}\mathbf{S}_i \right\rangle \right|  
\label{eqn:M_t formula2}
\end{eqnarray}
where $ \sum_{\gamma}$ denotes the sum over three sublattices. The order parameter defined by Eq. \ref{eqn:M_t formula2}  is calculated from cooling cycles for both three and six layer films is plotted as a function of temperature in  Fig.~\ref{fig:Mt}.
The transition temperatures deduced from these figures correspond approximately to those estimated from the specific heat peaks. Differences between the three and six layer films are attributed to the relative contributions of the surface {\it vs} interior layers. The plots of $M_t$  show that for three layer films the order parameter is not always saturated at zero temperature and that some of the points show over a much wider range of scatter around the transition temperature than the corresponding results for the six layer case. These features  are largely independent of the value of $D$. Such effects may be attributed to the presence of planes of atoms in which pairs of NN spins have been interchanged as discussed earlier in Section~\ref{GroundState} and described for the 3D case in Refs. ~\cite{hemmati2012} and ~\cite{leblanc2013} . In the absence of the cubic anisotropy (ie $K=0$) the ground state energy of the thin film, like the 3D fcc Kagome lattice, is left invariant by the presence of such defects. This give rise to the large degeneracy of the ground state spin configurations that characterises the kagome lattice in both two and three dimensions. The presence of a finite cubic anisotropy (ie $K>0$) lifts this degeneracy and serves to suppress the presence of such defects at low temperatures. Previous simulation studies in the case show that a value of $K=0.1$ is sufficient to remove such defects over the entire temperature range $0 < T< T_N$ \cite{hemmati2012}. 

In the context of the current model the effects of the anisotropy terms on these defects are somewhat more complicated. It can readily be shown that the axial surface anisotropy leaves the energy invariant under the interchange of NN spins along planes of atoms. This is consistent with the observation that the scatter in the order parameter near $T_N$ and the lack of saturation at the lowest temperature observed in the results presented in Figs. \ref{fig:Mt} is largely independent of the parameter $D$. The suppression of such features with increasing thickness reflect the fact that the energy of such defects, due to the cubic anisotropy, will be proportional to the number of interior layers. 

\section{Magnetization}

\begin{figure}[!htbp] 
		\centering
		\includegraphics[width=0.7\linewidth]{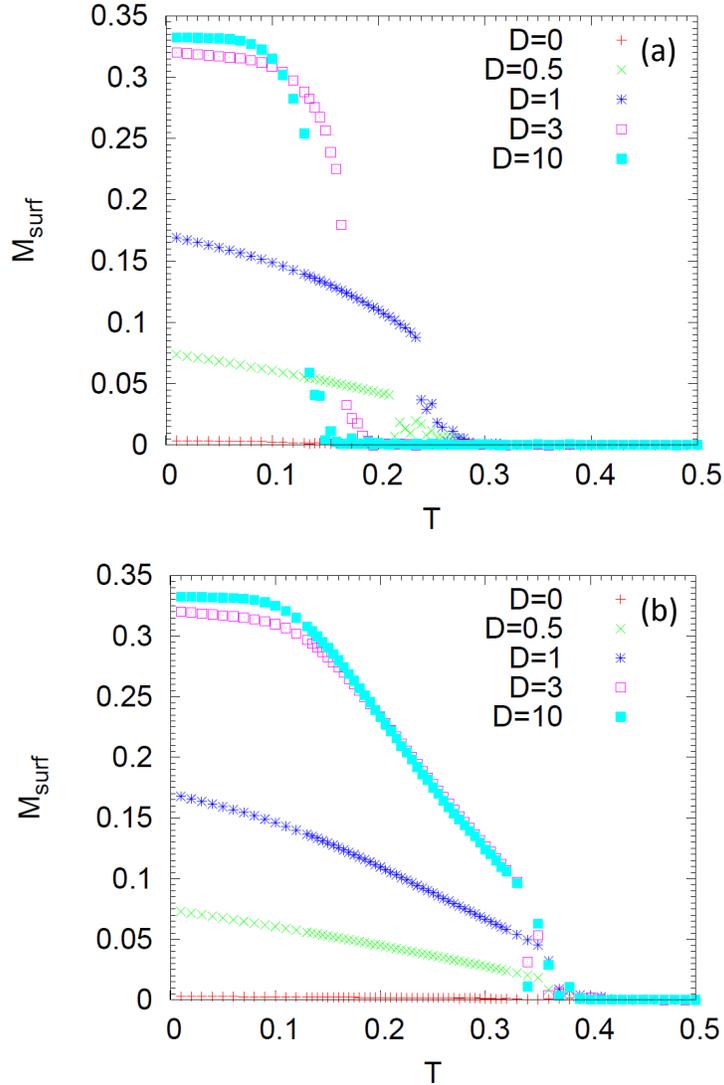}
\caption{Magnetization of the surface layers from (a) three layer and (b) six layer films. }
\label{fig:Mfsurface}		
\end{figure}

In Sec. III it was shown that the ground state for both three and six layer films is characterised  by ferrimagnetic surface layers with a saturated magnetization $M_\mathrm{surf} = 1/3$ per spin in the limit $D\to \infty$ with essentially antiferromagnetic interior layers. In Fig.~\ref{fig:Mfsurface} the surface magnetization per spin is plotted as a function temperature for several values of $D$. While the data around $T_N$ show a lot of scatter, the magnetization for both the three and six layer films appears to be significant over the temperature range $0 < T< T_N$. Projections of the magnetizations of the top layer onto the normal to the film ($z'$-axis, parallel to [111]) were also calculated and are nearly identical to the results of Fig.~\ref{fig:Mfsurface} indicating that the magnetization vector points nearly perpendicular to the film plane and that there is no preferrential direction (up or down) along the $z'$ axis.

The data in the six layer case also show, except close to $T_N$, that the surface magnetization increases monotonically with increasing $D$ for a given $T$ (consistent with the ground state results of Fig. 7). In the three layer case, the dependence of the surface magnetization shows a more complex dependence on $D$ and $T$ due to the fact that $T_N$ shows a strong dependence on $D$. However, it can be shown that, even for the three layer case, the surface magnetization increases with increasing $D$ for a constant value of $T/T_N(D)$. 

\begin{figure}[!htbp] 
		\centering
		\includegraphics[width=0.7\linewidth]{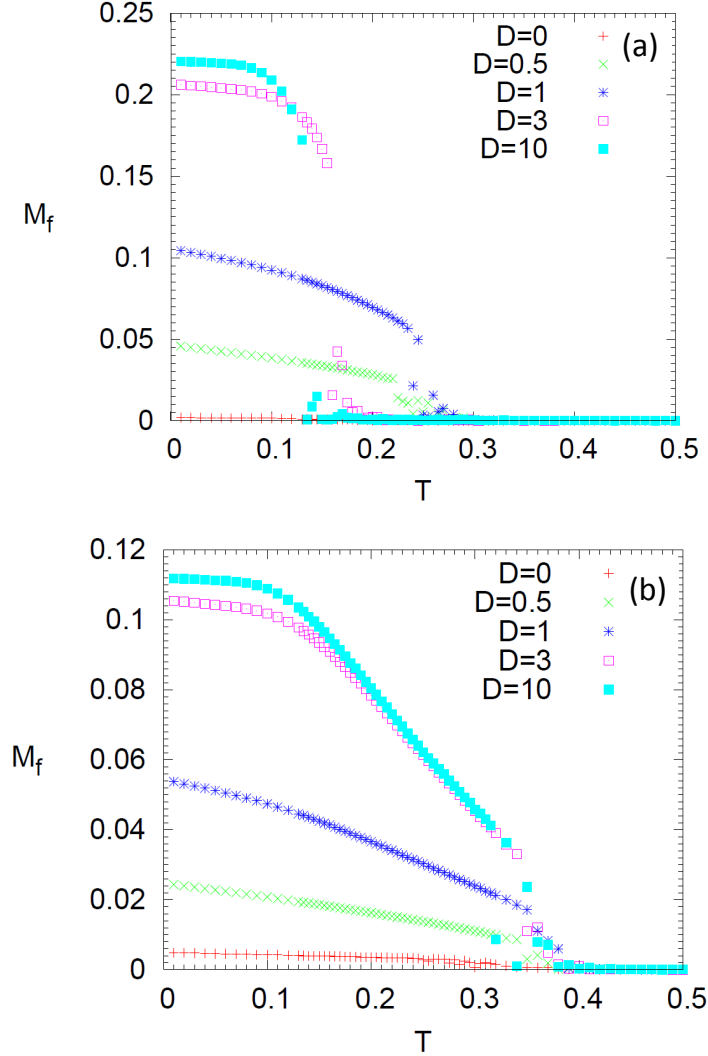}		
\caption{Magnetization $M_\mathrm{f}$  from (a) three layer and (b) six layer films.}
\label{fig:Mf}
\end{figure}

The magnetization of the interior layers was also calculated and found to be small, less than 0.03, over the same range of $D$ and $T$ values. This result is expected based on the ground state magnetization of the interior layers shown in Fig.~\ref{fig:GSmagnetization}. The antiferromagnetic character of the interior also is reflected in the plots of temperature dependence of the magnetization per spin of the entire film $M_\mathrm{f}$ shown in Fig.~\ref{fig:Mf}. Noting that $M_\mathrm{f}$ may be written as 
\begin{eqnarray}
M_\mathrm{f}&= M_\mathrm{int} + \frac{2}{L} \left(M_\mathrm{surf} -M_\mathrm{int}\right)
\end{eqnarray}
and assuming $M_\mathrm{int} \ll M_\mathrm{surf}$, we obtain the approximate scaling relation $M_\mathrm{f}\approx 2M_\mathrm{surf}/L$. 

We close this section by noting that the finite temperature results for six layer film, like the ground state calculations of the previous section, show many of the characteristics one would expect of a thick film, while the behaviour of the three layer film is dominated by the surface magnetization and its properties are strongly dependent on the value of $D$. While the magnetic properties of the three layer film are interesting, the properties of the six layer film are more representative of thicker films where the effects of the surface anisotropy are limited to the first few layers and do not significantly affect the intensive properties of the film. That the effects of the surface anisotropy, most notably the surface magnetization, do not appear to propagate beyond the first few layers is a significant result and represent a desirable property from the perspective of EB. 

\section{Conclusions}

We have extended previous MC studies of classical AF magnetic order in 2D and 3D kagome lattices to the case of (fcc) ABC stacked thin films.  A focus of the present work has been to examine the impact of axial surface ($D$) and bulk cubic ($K$) anisotropies on the spin configurations and degeneracies for three and six layer films.
It is shown that for the six layer case that the presence of the surface anisotropy leads to a relatively small change in the value of $T_N$ from its 3D value.  This is in contrast with the three layer case where $T_N$ vs $D$ exhibits a maximum near $D = 0.7$ with $T_N=0.3$ before decreasing to a nearly constant value of $T_N\approx 0.15$ for $D$ greater than about 4.  (In contrast, for the case of ferromagnetically exchange coupled spins,  $T_N$ increases monotonically with increasing $D$ \cite{leblanc2013b}.)  In addition, the magnetic structure of the interior layers is qualitatively similar to the $q=0$ spin structure of the bulk material described in ~\cite{leblanc2013}, while the spins at the surface layers have a component normal to the (111) plane  and order ferrimagnetically.  

Results for the order parameter $M_t$ as a function of temperature show switching between nearly degenerate kagome 120$^0$ spin states as in the bulk case, but only for the three layer film where surface effects are more pronounced. Switching is much less probable in the six layer case. This suggests that the magnetic structure of thicker films would be similar to the six layer films studied here, with values for $T_N$ close to that of the bulk material

One of the most striking features of the results is that a moderate to large value of the surface anisotropy parameter $D$ induces a ferrimagnetic arrangement of the spins on the surface of the film with a net magnetization directed perpendicular to the film. The results of these simulations imply that a perpendicular surface  anisotropy in IrMn$_3$ would induce a robust surface magnetization, that persists up to $T_N$,  while the interior of the film remains antiferromagnetic with a small net magnetization. This suggests a mechanism for EB that is relevant to the fcc kagome structure of IrMn$_3$. However, to what extent this accounts for the pinning mechanism in current spin valves is not yet clear. For example, the fact that the surface magnetization is perpendicular to the surface means that a simple exchange coupling between the IrMn$_3$ and the planar ferromagnetic Co layer would not produce exchange bias in the parallel field. However,  more complex coupling, e.g. long-range dipole interactions \cite{meloche2011,holden2015} or anti-symmetric exchange \cite{yanes2013}, could result in a coupling between an fcc kagome lattice and a planar ferromagnet.  It is also of interest to examine the influence of surface (Ir-Mn) disorder as well as vacancies \cite{biternas2014} on the magnetic domain structure believed to be relevant for the field-cooled-induced in-plane uniaxial effective anisotropy associated with the EB phenonmenon \cite{ogrady}.

\vspace{0.2in}
This work was supported by the Natural Science and Engineering Research Council (NSERC) of Canada and Compute Canada.
\vspace{0.2in}

{\bf Appendix} 
\vspace{0.2in}

An analsis of the ground state in the case $0<D<<J$, $0<K<<J$ is outlined here where perturbation theory is used. Consider first the case with $D=0,\; K=0$ when spins lie in the $(11\bar{1})$ plane \cite{leblanc2013}.  We note that the reduced number of NNs at surfaces does not change the spin structure from the bulk case so that we take
\begin{subequations}
\begin{align}
\mathbf{S}_1&=(-\sqrt{\frac{2}{3}},\frac{1}{\sqrt{6}},-\frac{1}{\sqrt{6}}),   \\  
\mathbf{S}_2&=(\frac{1}{\sqrt{6}},-\sqrt{\frac{2}{3}},-\frac{1}{\sqrt{6}}),   \\  
\mathbf{S}_3&=(\frac{1}{\sqrt{6}},\frac{1}{\sqrt{6}},\sqrt{\frac{2}{3}}).
\label{oth Order}
\end{align}
\end{subequations}
It is convenient to work in a polar coordinate system, since while minimizing, the condition ${S_{ix}}^2+{S_{iy}}^2+{S_{iz}}^2=1$ is fulfilled automatically. Then, spins $\mathbf{S}_1$, $\mathbf{S}_2$, $\mathbf{M}_1$, and $\mathbf{M}_2$  are described with a polar and an azimutal angles $(\theta_1, \phi_1)$, $(\theta_1, \frac{\pi}{2}-\phi_1)$, $(\alpha_1, \beta_1)$ and $(\alpha_1, \frac{\pi}{2}-\beta_1)$, respectively, and spins $\mathbf{S}_3$ and $\mathbf{M}_3$ are described with angles $(\theta_3, \frac{\pi}{4})$ and  $(\alpha_3, \frac{\pi}{4})$.

To account for the effects of small $K$ and $D$, we write the angles which minimize the energy as an expansion in the form

\begin{eqnarray}
\theta_1=\theta^{(0)}_1+\theta^{(1)}_1  \;\;\;\;&    \phi_1=\phi^{(0)}_1+\phi^{(1)}_1  \;\;\;\; &   \theta_3=\theta^{(0)}_3+\theta^{(1)}_3  \nonumber \\
\alpha_1=\alpha^{(0)}_1+\alpha^{(1)}_1  \;\;\;\;&    \beta_1=\beta^{(0)}_1+\beta^{(1)}_1   \;\;\;\;&   \alpha_3=\alpha^{(0)}_3+\alpha^{(1)}_3,   
\end{eqnarray}
where the superscript (0) corresponds to the case $K= D = 0$, and
\begin{eqnarray}
\theta^{(0)}_1=\arccos\frac{1}{\sqrt{6}}; &   \phi^{(0)}_1=-\arctan\frac{1}{2}; &   \theta^{(0)}_3=\arccos\sqrt{\frac{2}{3}},   \nonumber \\
\alpha^{(0)}_1=\arccos\frac{1}{\sqrt{6}}; &   \beta^{(0)}_1=-\arctan\frac{1}{2};  &   \alpha^{(0)}_3=\arccos\sqrt{\frac{2}{3}}.   
\end{eqnarray}
After expansion of the energy to second order, minimization requires a solution to the following matrix equation:
\begin{eqnarray}
\begingroup\makeatletter\def\f@size{6}\check@mathfonts
\hspace*{-5.1cm}
 \left(
\begin{array}{cccccc}
\frac{16D}{405}-\frac{88J}{45} & -\frac{8D}{27\sqrt{5}}+\frac{4J}{9\sqrt{5}} &  -\frac{8 \sqrt{2} J}{9\sqrt{5}} &  -\frac{14 J}{45} &
 \frac{2 J}{ 9 \sqrt{5}} &  -\frac{4}{9}\sqrt{\frac{2}{5}}J \\
-\frac{8 D}{27 \sqrt{5}} + \frac{4 J}{9 \sqrt{5}} &  -\frac{28 D}{81} - \frac{46 J}{27} &  \frac{4 \sqrt{2} J}{9} &  \frac{2 J}{ 9 \sqrt{5}} &
-\frac{8 J}{27} &  \frac{2 \sqrt{2}J}{9} \\
 -\frac{8 \sqrt{2} J}{9\sqrt{5}} &  \frac{4 \sqrt{2} J}{9} &  \frac{28 D}{81} - \frac{2 J}{3} &  -\frac{4}{9} \sqrt{\frac{2}{5}} J &
 \frac{2 \sqrt{2} J}{9} &  0 \\
-\frac{14 J}{45} &  \frac{2 J}{9 \sqrt{5}} & -\frac{4}{9} \sqrt{\frac{2}{5}} J &  -\frac{6 J}{5} + \frac{32 K}{135} &
\frac{2 J}{9 \sqrt{5}} + \frac{8 K1}{27 \sqrt{5}} &  -\frac{4 \sqrt{2} J}{9\sqrt{5}} \\
\frac{2 J}{9 \sqrt{5}} &  -\frac{8 J}{27} &  \frac{2 \sqrt{2} J}{9} &  \frac{2 J}{9 \sqrt{5}} + \frac{8 K}{27 \sqrt{5}} &
 -\frac{28 J}{27} + \frac{2 K}{ 9} &  \frac{2 \sqrt{2} J}{9} \\
-\frac{4}{9} \sqrt{\frac{2}{5}} J &  \frac{2 \sqrt{2} J}{9} & 0 &  -\frac{4 \sqrt{2} J}{9\sqrt{5}} & 
 \frac{2 \sqrt{2} J}{9} &  -\frac{4 J}{9} + \frac{2 K}{27} 
\end{array} \right)
\left ( 
\begin{array}{c}
\theta^{(1)}_1\\
\phi^{(1)}_1\\
\theta^{(1)}_3\\
\alpha^{(1)}_1\\
\beta^{(1)}_1\\
\alpha^{(1)}_3
\end{array} \right )
=
\left ( 
\begin{array}{c}
\frac{32 D}{81 \sqrt{5}}\\
\frac{8 D}{27}\\
\frac{8 \sqrt{2} D}{81}\\
-\frac{8 K}{27 \sqrt{5}}\\
\frac{4 K}{27}\\
-\frac{2 \sqrt{2} K}{27}
\end{array} \right ).
\endgroup
\end{eqnarray}

\vspace{0.5cm}

For small values of $D$, the solution obtained by this approximation is in excellent agreement with a direct numerical minimization  of the energy Eq.~(\ref{eqn:energy_per_spin}) (using Wolfram Mathematica).


\vspace{0.2in}

\begin{thebibliography}{10}

\bibitem{gorohovsky2015}
Gorohovsky B, Pereira R G  and  Sela E 2015
\newblock {\em Phys. Rev. B} \textbf{91} 245139 

\bibitem{carrasquilla2015}
Carrasquilla J, Hao Z H, Melko R G 2015
\newblock {\em Nature Commun.}  \textbf{6} 7421 

\bibitem{chalker1992}
Chalker J T, Holdsworth P C W and Shender E F 1992
\newblock {\em Phys. Rev. Lett.}  \textbf{68} 855 

\bibitem{harris1992}
Harris A B, Kallin C  and Berlinsky A J 1992
\newblock {\em Phys. Rev. B}  \textbf{45} 2899 

\bibitem{schnabel2012}
 Schnabel S and  Landau D P 2012
\newblock {\em Phys. Rev. B}  \textbf{86} 014413 

\bibitem{mendels2011}
Mendels P and  Wills A S 2011
\newblock {\em Introduction to Frustrated Magnetism}, eds Lacroix  C, Mendels  P and Mila F  {\em
  Springer Series in Solid-State Sciences} \textbf{164} ( Heidelberg: Springer)

\bibitem{zhitomirsky2008}
Zhitomirsky M E 2008
\newblock {\em Phys. Rev. B}  \textbf{78} 094423 

\bibitem{tomeno1999}
Tomeno I, Fuke H N, Iwasaki H, Sahashi M andTsunoda Y 1999
\newblock {\em J. Appl. Phys.}  \textbf{86}  3853 

\bibitem{szunyogh2009}
Szunyogh L, Lazarovits B, Udvardi L,  Jackson J  and Nowak U 2009
\newblock {\em Phys. Rev. B} \textbf{79} 020403 

\bibitem{hemmati2012}
Hemmati V, Plumer M L,  Whitehead J P  and Southern B W 2012
\newblock {\em Phys. Rev. B} \textbf{86} 104419 

\bibitem{leblanc2013}
LeBlanc M D,  Plumer M L,  Whitehead J P  and Southern B W 2013
\newblock {\em Phys. Rev. B} \textbf{88} 094406 

\bibitem{leblanc2014}
LeBlanc M D,  Southern B W, Plumer M L and  Whitehead J P  2014
\newblock {\em Phys. Rev. B}  \textbf{90} 144403 

\bibitem{jensen2006}
Jensen P J and Bennemann K H 2006
\newblock {\em Surf. Sci. Reps.}  \textbf{61} 129 

\bibitem{bland2008}
Vaz C A F, Bland J A C  and G.~Lauhoff G 2008
\newblock {\em Rep. Prog. Phys.}  \textbf{71} 056501 

\bibitem{popov2008}
Popov A P, Skorodumov N V and Erikson O 2008
\newblock {\em Phys. Rev. B}  \textbf{77} 014415 

\bibitem{binder1992}
Binder K 1992
\newblock {\em Annu. Rev. Phys. Chem.}  \textbf{43} 33 

\bibitem{diep2015}
Diep H T 2015
\newblock {\em Phys. Rev. B}  \textbf{91} 014436 

\bibitem{haraldsen2010}
Haraldsen J T and Fishman R S 2010
\newblock {\em Phys. Rev. B}  \textbf{81} 020404(R) 

\bibitem{wilson2014}
Wilson M N, Butenko A B,  Bogdanov A N and  Monchesky T L 2014
\newblock {\em Phys. Rev. B}  \textbf{89} 094411 

\bibitem{langridge2014}
Langridge S, Watson G M, Gibbs D, Betouras J J, Gidopoulos N I,
Pollmann F, Long M W, Vettier C and Lander G H 2014
\newblock {\em Phys. Rev. Lett.}  \textbf{112} 167201 

\bibitem{wang2013}
Wang B-Y, Hong J-Y,  Yang K-H O, Chan Y-L, Wei D-H, Lin H-J and M.-T. Lin M-T 2013 
\newblock {\em Phys. Rev. Lett.}  \textbf{110} 117203 

\bibitem{tsunoda2009}
Tsunoda M, Takahashi H, and Takahashi M 2009
\newblock {\em IEEE Trans Magn.}  \textbf{45} 3877 

\bibitem{tsunoda2010}
Tsunoda M, Takahashi H, Nakamura T, Mitsumata C, Isogami S, and Takahashi M 2010
\newblock {\em Appl. Phys. Lett.}  \textbf{97} 072501 

\bibitem{kohn2013}
Kohn A, Kova A, Fan R,  McIntyre G J, Ward R C C and  Goff J P 2013
\newblock {\em Scientific Reports}  \textbf{45} 2899 

\bibitem{szunyogh2011}
Szunyogh L, Udvardi L, Jackson J, Nowak U  and Chantrell R 2011
\newblock {\em Phys. Rev. B} \textbf{83} 024401 

\bibitem{yanes2013}
Yanes R, Jackson R, Udvardi L, Szunyogh L  and Nowak U 2013
\newblock {\em Phys. Rev. Lett.}  \textbf{111} 217202 

\bibitem{walker1980}
Walker L R and Walstedt R E 1980
\newblock {\em Phys. Rev. B} \textbf{22} 3816  

\bibitem{leblanc2013b}
LeBlanc M D,  Whitehead J P and Plumer M L 2013
\newblock {\em J. Phys.: Condens. Matter}  \textbf{25} 196004 

\bibitem{meloche2011}
Meloche E,  Mercer J I,  Whitehead J P,  Nguyen T M and Plumer M L 2011
\newblock {\em Phys. Rev. B}  \textbf{83} 174425 

\bibitem{holden2015}
Holden M S, Plumer M L, Saika-Voivod I and  Southern B W 2015
\newblock {\em Phys. Rev. B}  \textbf{91} 224425 

\bibitem{biternas2014}
Biternas A G, Chantrell R W and Nowak U 2014
\newblock {\em Phys. Rev. B}, \textbf{89} 184405 

\bibitem{ogrady}
O'Grady K, Fernandez-Outon L E and Vallejo-Fernande G 2010
\newblock {\em J. Magn. Magn. Mat.}, \textbf{322} 883

\end{thebibliography}

\end{document}